\documentclass[12pt]{iopart}
\usepackage{graphicx}
\input amssym.def \input amssym

\begin{document}

\title[]{On group theory for quantum gates \\and quantum coherence}

\author{Michel Planat$^{\dag}$ and Philippe Jorrand$^{\ddag}$ }

\address{$^{\dag}$ Institut FEMTO-ST, CNRS, 32 Avenue de
l'Observatoire,\\ F-25044 Besan\c con, France. }



\address{$^{\ddag}$ Laboratoire d'Informatique de Grenoble, 110 av. de la Chimie, \\Domaine Universitaire, BP 53, 38041 Grenoble cedex 9, France.}

\begin{abstract}
Finite group extensions offer a natural language to quantum computing. In a nutshell, one roughly describes the action of a quantum computer as consisting of two finite groups of gates: error gates from the general Pauli group $\mathcal{P}$ and stabilizing gates within an extension group $\mathcal{C}$. In this paper one explores the nice adequacy between group theoretical concepts such as commutators, normal subgroups, group of automorphisms, short exact sequences, wreath products... and the coherent quantum computational primitives. The structure of the single qubit and two-qubit Clifford groups is analyzed in detail. As a byproduct, one discovers that $M_{20}$, the smallest perfect group for which the commutator subgroup departs from the set of commutators, underlies quantum coherence of the two-qubit system. One recovers similar results by looking at the automorphisms of a complete set of mutually unbiased bases.
\end{abstract}


\maketitle

\noindent
\hrulefill
\section{Introduction}

Currently quantum computing is a very active and respectable area of research at the interface of the three pillars: quantum physics, mathematics and  computer science. If large-scale quantum computers can be built, they will be able to solve certain problems, such as quantum factoring, quantum search or the graph isomorphism problem, in a very efficient way when compared to classical computing. However, one of the main drawbacks of quantum computing is its extreme sensitivity to the classical environment. The irreversible formation of quantum correlations of a system with the surrounding world induces the decoherence of quantum preparations. To overcome this limitation, many designs have been proposed for correcting the unavoidable errors, or for preventing them to occur. Since the inception of the field,  fault-tolerant procedures such as universal bases of gates \cite{Boykin}, quantum codes \cite{Go05} or quantum teleportation based protocols \cite{Gottesman1}  have been proposed. Other approaches relate to topological quantum computation \cite{Kitaev97,Raussen08}, decoherence free subspaces \cite{Zanardi} or are based on sequences of measurements \cite{Jorrand}. 

Despite the number of seemingly different proposals some of them are related: there is a close relation between the \lq\lq oldfashioned" quantum gate circuitry, fault tolerant quantum codes and measurements, already apparent in the stabilizer formalism \cite{Go98, Nielsen}. It was shown that a few building block gates are enough to simulate any unitary evolution \cite{Go05} and a few minimal resources are required for measurement-only quantum computation \cite{Perdrix}. This paper explores the fresh view that the geometry of commutation relations \cite{Pauligraphs}-\cite{Saniga} between the error operators, their corresponding group of symmetries (i.e. the automorphisms), and the splitting of the stabilizer group in terms of maximal normal subgroups \cite{Mathieu}, sustain the explanation of quantum (de)coherence. Although the approach is performed for a reduced number of qubits, novel pieces of the puzzle appear such as perfect groups with special group theoretical or geometrical properties, and new links are established, such as the relevance of mutually unbiased bases to quantum coherence, or the embedding of quantum topological concepts within the Clifford group. Several recent papers concern closely related topics, see for example Refs \cite{Clark07}-\cite{Grassl}.

Following an outline of useful group theoretical concepts in Sec \ref{intro}, the structure of one and two-qubit Clifford groups is unraveled in Sec \ref{Clifford} in terms of split short exact sequences, which makes use of permutation groups acting on five or six letters. Calculations are performed using GAP \cite{GAP} and MAGMA \cite{MAGMA}.

\section{An outline of group commutators, group extensions and groups of automorphisms}
\label{intro}

For an introduction to group theory one may use the on-line Ref \cite{Milne}.
A subgroup $N$ of a group $G$ is called a normal subgroup if it is invariant under conjugation: that is, for each $n$ in $N$ and each $g$ in $G$, the conjugate element $g n g^{-1}$ still belongs to $N$. In particular, the center $Z(G)$ of a group $G$ (the set of all elements in $G$ which commute with each element of $G$) is a normal subgroup of $G$. The group $\tilde{G}=G/Z(G)$ is called the central quotient of $G$. A second important example of a normal subgroup of $G$  is provided by the subgroup $G'$ of commutators (also called the derived subgroup of $G$). It is defined as the subgroup generated by all the commutators $[g,h]=ghg^{-1}h^{-1}$ of elements of $G$. The quotient group $H^{\mbox{ab}}=G/G'$  is an abelian group called the abelianization of $G$ and corresponds to its first homology group. The set $K(G)$ of all commutators of a group $G$ may depart from $G'$ \cite{kappe}. 

Our third example relates to group extensions.
Let $\mathcal{P}$ and $\mathcal{C}$ be two groups such that $\mathcal{P}$ is normal subgroup of $\mathcal{C}$. The group $\mathcal{C}$ is an extension of $\mathcal{P}$ by $H$ if there exists a short exact sequence of groups
\begin{equation}
1 \rightarrow \mathcal{P} \stackrel{f_1}{\rightarrow} \mathcal{C} \stackrel{f_2}\rightarrow H \rightarrow 1,
\nonumber
\end{equation}
in which $1$ is the trivial (single element) group.

The above definition can be reformulated as follows

(i) $\mathcal{P}$ is isomorphic to a normal subgroup $N$ of $\mathcal{C}$,

(ii) $H$ is isomorphic to the quotient group $\mathcal{C}/N$.

Because in an exact sequence the image of $f_1$ is equal to the kernel of $f_2$, then the map $f_1$ is injective and $f_2$ is surjective. 

* Given any groups $\mathcal{P}$ and $H$ the direct product of $\mathcal{P}$ and $H$ is an extension of $\mathcal{P}$ by $H$.

* The semidirect product $\mathcal{P} \rtimes H$ of $\mathcal{P}$ and $H$ is defined as follows.  The group $\mathcal{C}$ is an extension of $\mathcal{P}$ by $H$ (one identifies $\mathcal{P}$ with  a normal subgroup of $\mathcal{C}$) and 

(i) $H$ is isomorphic to a subgroup of $\mathcal{C}$,

(ii) $\mathcal{C}$=$\mathcal{P} H$ and

(iii) $\mathcal{P}\cap H=\left\langle 1\right\rangle$.

One says that the short exact sequence splits.   

The wreath product $M \wr H$ of a group $M$ with a permutation group $H$ acting on $n$ points is the semidirect product of the normal subgroup $M^n$ with the group $H$, which acts on $M^n$ by permuting its components.

* Let $G=\mathcal{Z}_2 \wr A_5$, in which $A_5$ is the alternating group on five letters, then $G'$ is a perfect group with order $960$ and  one has $G' \ne K(G)$. Let $H=Z_2^5 \rtimes A_5$, one can think of $A_5$ having a wreath action on $Z_2^5$. The group  $G'=\tilde{H}=M_{20}$  \cite{Brauer} is the smallest perfect group having its commutator subgroup distinct from the set of the commutators \cite{kappe}. One easily checks that $M_{20}$ also corresponds to the derived subgroup $W'$ of the Weyl group (also called hyperoctahedral group) $W=\mathcal{Z}_2 \wr S_5$ for the Lie algebra of type $B_5$. For a quantum version, see \cite{Banica}.  

\subsection*{Group of automorphisms}

Given the group operation $\ast$ of a group $G$, a group endomorphism is a function $\phi$ from $G$ to itself such that $\phi(g_1 \ast g_2)= \phi(g_1) \ast \phi(g_2)$, for all $g_1,g_2 \in G$. If it is bijective, it is called an automorphism. An automorphism of $G$ that is induced by conjugation of some $g\in G$ is called inner. Otherwise it is called an outer automorphism. Under composition the set of all automorphisms defines a group denoted $\mbox{Aut}(G)$. The inner automorphisms form a normal subgroup $\mbox{Inn}(G)$ of $\mbox{Aut}(G)$, that is isomorphic to the cental quotient of $G$. The quotient $\mbox{Out}(G)=\mbox{Aut}(G)/\mbox{Inn}(G)$ is called the outer automorphism group.

\section{Quantum computing and the Clifford group}
\label{Clifford}

Compared to group theory, the science of quantum computing is in its infancy \cite{Nielsen}. In quantum codes and in quantum computing, one is interested in preventing or correcting errors that may affect one or many physical qubits \cite{Calder}-\cite{Kla2}. A frequently used error group is the general Pauli group $\mathcal{P}_n$. It consists of tensor products of the Pauli matrices \cite{Pauligraphs}
\begin{equation}
\sigma_x =\left(\begin{array}{cc} 0 & 1 \\1	 & 0\end{array}\right),~~\sigma_z=\left(\begin{array}{cc} 1 & 0 \\0	 & -1\end{array}\right), ~~\sigma_y=i\sigma_x\sigma_z,
\nonumber
\end{equation}  
and the unity matrix $\sigma_0$. Pauli matrices generate the single qubit Pauli group $\mathcal{P}_1$ of order $16$ and center $Z(\mathcal{P}_1)=\{\pm 1,\pm i\}$. More generally the $n$-qubit Pauli group $\mathcal{P}_n$, of order $4^{n+1}$, is generated by the tensor product of $n$ Pauli matrices. 

Let us assume a quantum computer in a state $\left|\psi\right\rangle$, and apply to it an error $g$ belonging to the Pauli group $\mathcal{P}$ so that the new state of the computer is $g\left|\psi\right\rangle $. One allows unitary evolutions $U$ so that the new state evolves as $ Ug\left|\psi\right\rangle=UgU^{\dag} U\left|\psi\right\rangle$. For stabilizing the error within the Pauli group $\mathcal{P}$, one requires that $UgU^{\dag} \in \mathcal{P}$. The set of operators leaving $\mathcal{P}$ invariant under conjugation is the normalizer $\mathcal{C}$ in the unitary group $U$, also known as the Clifford group \cite{Go98}-\cite{Calder}\footnote{The Clifford group (also known as the Jacobi group) was introduced in the context of quantum stabilizer codes by D Gottesman. It doesn't explicitely refer to Clifford algebras in which the Clifford group means \lq\lq the set of invertible elements  in the Clifford algebra that stabilize under twisted conjugation". In the context of a $n$-qubit system, a Clifford algebra may be obtained by selecting a set of mutually anti-commuting observables as for the Dirac relativistic equation.}.  Within a unitary group one has the equality $U^{\dag}=U^{-1}$. As a result, the group $\mathcal{P}$ is a normal subgroup of $\mathcal{C}$ and one  vectors and one may use the powerful formalism of group extensions to report on it. Additionaly some subgroups of $\mathcal{C}$, which have the error group $\mathcal{P}$ as a normal subgroup, will play a role for displaying the quantum coherence.      

The Clifford group, stabilizing the (error) Pauli group $\mathcal{P}_n$ on n-qubits, will be denoted $\mathcal{C}_n$. One learned from Gottesman-Knill theorem that the Hadamard  gate $H=1/\sqrt{2}\left(\begin{array}{cc} 1 & 1 \\1	 & -1\end{array}\right)$ and the phase gate $P=\mbox{Diag}(1,i)$ are in the one-qubit Clifford group $\mathcal{C}_1$, and that the controlled-$Z$ gate $CZ=\mbox{Diag}(1,1,1,-1)$ is in the two-qubit Clifford group $\mathcal{C}_2$. Any gate in $\mathcal{C}_n$ may be generated from the application of gates from $\mathcal{C}_1$ and $\mathcal{C}_2$ \cite{Go98,Clark07}. Clifford group $\mathcal{C}_n$ on $n$-qubits has order $|\mathcal{C}_n|=2^{n^2+2n+3}\prod_{j=1}^n 4^j-1$ \cite{Calder}.

Below we will concentrate on the properties of the Clifford group related to one and two qubits, using the group theoretical package GAP4 \cite{GAP}. Generation of the gates will be ensured by the use of cyclotomic numbers, as described in Sec 18 of the GAP4 reference manual. For example, the elements $1$, $-1$, $i$ and $2^{1/2}$ will be modelled as the roots of unity $E(1)$, $E(2)$, $E(4)$ and as $ER(2)$, respectively.

\subsection{The Clifford group on a single qubit}

The one-qubit Clifford group is generated bu $H$ and $P$ as $\mathcal{C}_1=\left\langle H,P \right\rangle$. It has order $|\mathcal{C}_1|=192$, its center is $Z(\mathcal{C}_1)=\mathcal{Z}_8$ and the derived subgroup $\mathcal{C}'_1$ equals the special linear group $SL(2,3)$. The central quotient is $\tilde{\mathcal{C}_1}=S_4$ and one obtains the abelianization as the direct product $\mathcal{C}_1^{\mbox{ab}}=\mathcal{Z}_4 \times \mathcal{Z}_2$.  

Using the method described in Sec \ref{intro} two split extensions follow. The first one is attached to $\mathcal{C}_1'=SL(2,3)$ as follows  
\begin{equation}
1 \rightarrow SL(2,3) {\rightarrow} ~\mathcal{C}_1\rightarrow  \mathcal{Z}_2 \times  \mathcal{Z}_3  \rightarrow 1.
\nonumber
\end{equation}
The second one is attached to the Pauli group 
\begin{equation}
1 \rightarrow \mathcal{P}_1 {\rightarrow}~ \mathcal{C}_1\rightarrow  D_{12}  \rightarrow 1,
\nonumber
\end{equation}
in which $D_{12}=\mathcal{Z}_2 \times S_3$ is the dihedral symmetry group of a regular hexagon.

\subsection{The Clifford group on two qubits}
\label{quadrangle}

The two-qubit Pauli group may be generated as \\ $\mathcal{P}_2=\left\langle\sigma_x \otimes \sigma_x,\sigma_z \otimes \sigma_z,\sigma_x \otimes \sigma_y,\sigma_y \otimes \sigma_z,\sigma_z \otimes \sigma_x \right\rangle$. It is of order $64$ and has center $\mathcal{Z}(\mathcal{P}_2)=\mathcal{Z}(\mathcal{P}_1)$.  The two-qubit Clifford group, of order 92160, may be generated from $H$, $P$ and $CZ$ as $\mathcal{C}_2=\left\langle H \otimes H, H \otimes P, CZ\right\rangle$. Its center is $Z(\mathcal{C}_2)=Z(\mathcal{C}_1)$ and the central quotient $\tilde{\mathcal{C}_2}$ is found to satisfy the exact sequence
\begin{equation}
1 \rightarrow U_6 \rightarrow \tilde{\mathcal{C}_2} \rightarrow  \mathcal{Z}_2  \rightarrow 1,
\nonumber
\end{equation}
in which we introduced the notation $U_6=\tilde{\mathcal{C}}_2'=\mathcal{Z}_2^{\times 4} \rtimes A_6$. Another important relationship is $U_6=\mbox{Aut}(\mathcal{P}_2)'$, i.e. $U_6$ encodes the commutators of the Pauli group automorphisms.   It turns out that the group $\tilde{\mathcal{C}_2}$ only contains two normal subgroups $\mathcal{Z}_2^{\times 4}$ and $U_6$. The group $U_6$, of order $5760$, is a perfect group. It can be seen as a parent of the six element alternating group $A_6$. Its outer automorphism group $\mbox{Out}(U_6)$ is the same, equal to the Klein group $\mathcal{Z}_2 \times \mathcal{Z}_2$.

The group $U_6$ is an important maximal subgroup of several sporadic groups. The group of smallest size where it appears is the Mathieu group $M_{22}$. 
Mathieu groups are sporadic simple groups, so that $U_6$ is not normal in $M_{22}$. It appears in the context of a subgeometry of $M_{22}$ known as an {\it hexad}. 
Let us recall the definition of Steiner systems. A Steiner system $S(a,b,c)$ with parameters $a$, $b$, $c$, is a $c$-element set together with a set of $b$-element subsets of $S$ (called {\it blocks}) with the property that each $a$-element subset of $S$ is contained in exactly one block. A finite projective plane of order q, with the lines as blocks, is an $S(2, q+1, q^2+q+1)$, because it has $q^2+q+1$ points, each line passes through $q+1$ points, and each pair of distinct points lies on exactly one line. Any large Mathieu group can be defined as the automorphism (symmetry) group of a Steiner system \cite{Wilson}. The group $M_{22}$ stabilizes the Steiner system $S(3,6,22)$ comprising $22$ points and $6$ blocks, each set of $3$ points being contained exactly in one block\footnote{There exists up to equivalence a unique S(5,8,24) Steiner system called a Witt geometry. The group $M(24)$ is the automorphism group of this Steiner system, that is, the set of permutations which map every block to some other block. The subgroups $M(23)$ and $M(22)$ are defined to be the stabilizers of a single point and two points respectively.}. Any block in $S(3,6,22)$ is a Mathieu hexad, i.e. it is stabilized by the {\it general} alternating group $U_6$. 

There is a relationship between the two-qubit Clifford and Pauli groups
\begin{equation}
\mathcal{C}_2 / \mathcal{P}_2= \mathcal{Z}_2 \times S_6, 
\nonumber
\end{equation}
which features the important role of the six-letter symmetric group $S_6$. The latter governs the Pauli graph attached to the two-qubit system, being the automorphism group of generalized quadrangle of order two $W(2)$ \cite{Pauligraphs}. The group $S_6$ is peculiar among the symmetric permutation groups as having an outer automorphism group $\mathcal{Z}_2$.


\subsection{Quantum coherence within the two-qubit system}
\label{coherence}

Topological quantum computing based on anyons has been proposed as way of encoding quantum bits in nonlocal observables that are immune of decoherence \cite{Kitaev97,Preskill97}. The basic idea is to use pairs $\left|v,v^{-1}\right\rangle$ of \lq\lq magnetic fluxes" playing the roles of the qubits and permuting them within some large enough nonabelian finite group $G$ such as $A_5$. The \lq\lq magnetic flux" carried by the (anyonic) quantum particle is labeled by an element of $G$, and \lq\lq electric charges" are labeled by irreducible representation of $G$ \cite{Ogburn}.

The exchange within $G$ modifies the quantum numbers of the fluxons according to the fundamental logical operation
\begin{equation}
 \left|v_1,v_2\right\rangle \rightarrow \left|v_2,v_2^{-1} v_1 v_2\right\rangle,
\nonumber
\end{equation}
a form of Aharonov-Bohm interactions, which is nontrivial in a nonabelian group. This process can be shown to produce universal quantum computation. It is intimately related to topological entanglement, the braid group and unitary solutions of the Yang-Baxter equation \cite{Kauf}
\begin{equation}
(R \otimes I)(I \otimes R)(R \otimes I)=(I \otimes R)(R \otimes I)(I \otimes R),\nonumber
\end{equation}
in which $I$ denotes the identity transformation and the operator $R$: $V \otimes V \rightarrow V \otimes V$ acts on the tensor product of the bidimensional vector space $V$. One elegant unitary solution of the Yang-Baxter equation is a universal quantum gate known as the Bell basis change matrix
\begin{equation}
R=1/\sqrt{2}\left(\begin{array}{cccc} 1 & 0 & 0 & 1 \\0 & 1	& -1 & 0 \\ 0 & 1 & 1 & 0 \\-1 & 0 & 0 & 1\\ \end{array}\right)\nonumber.
\end{equation}
It is straightforward to see two-qubit topological quantum computing as another group extension of the Pauli group. One may introduce a subgroup of the Clifford group, of order $15360$, that we denote the Bell group as follows
\begin{equation}
\mathcal{B}_2=\left\langle H \otimes H, H \otimes P, R\right\rangle.
\end{equation}
The Bell group has center $\mathcal{Z}_8$ and its central quotient only contains two normal subgroups $\mathcal{Z}_2^{\times 4}$ and $M_{20}=\mathcal{Z}_2^{\times 4} \rtimes A_5$.
The group $M_{20}$ was already quoted in Sec \ref{intro} as being the smallest perfect group having the set of commutators departing from the commutator subgroup.
The relationship between the Bell and Pauli groups
\begin{equation}
\mathcal{B}_2 / \mathcal{P}_2= \mathcal{Z}_2 \times S_5 
\nonumber
\end{equation}
displays the important role of the five letters symmetric group $S_5$. At this point, it may be useful to mention that $A_5$ is the automorphism group of the icosahedron. Icosahedral symmetry and quantum coherence seems to be related in recent fullerene experiments \cite{Benjamin}.

\subsection{Quantum coherence within mutually unbiased bases}

To our knowledge, the relationship between mutually unbiased bases (MUBs) of the Pauli group and the Clifford group has not yet been established. Two orthonormal bases are said to be mutually unbiased if each common state of one basis gives rise to the same probability distribution when measured with respect to the other basis. For prime power dimensions $p^m$, complete sets of MUBs have cardinality $p^m+1$ and can be determined using different techniques such as the additive characters over a Galois field \cite{PlanatMUBs} \footnote{Power of prime dimensions also play a pivotal role in the number theoretical approach of $1/f$ noise developed by one of us \cite{noise1,noise2}.}. In composite dimensions, MUBs strongly rely on projective lines over finite rings \cite{PlanatJPhysA}. In addition, the continuous variable case was addressed recently \cite{Weigert}.

Commuting/non-commuting relations between the Pauli operators of the two-qubit system have been determined \cite{Pauligraphs}. The Pauli graph admits several decompositions: one of them is based on its minimum vertex cover (the Petersen graph) and a maximal independant set (of size five). If one uses a geometrical representation, operators correspond to the points of the geometry, maximal sets of mutually commuting operators, i.e. MUBs, correspond to the lines of the geometry, and a complete set of MUBs corresponds to an ovoid (the maximum number of mutually disjoint lines). The geometry of the two-qubit system is the smallest non-trivial generalized quadrangle. Due to the perfect duality between the fifteen points and fifteen lines of the quadrangle, the cardinality of a maximal independant set and the one of an ovoid is the same.

These graph theoretical and geometrical features of MUBs have a group theoretical counterpart that one may find in the group of automorphisms attached to a maximal independant set. Let us denote $m_i$ ($i=1..5$) the elements of such a maximal set, one may form groups of increasing size $g_2=\left\langle m_1,m_2\right\rangle$, ... $g_4=\left\langle m_1, m_2,m_3, m_4\right\rangle$. ($g_1$ is the trivial group and $g_5=g_4$). The groups $g_i$ and the corresponding groups of automorphisms $\mbox{Aut}(g_i)$ are identified in Table 1. One readily observes that the group of automorphisms of the selected maximal independant set/ovoid of the two-qubit system is isomorphic to the wreath product $\mathcal{Z}_2\wr A_5$ encountered in topological quantum computing. One concludes that some symmetries in a complete set of MUBs also provide a signature of quantum coherence. Let us mention that the hyperoctahedral group  $\mathcal{Z}_2 \wr S_5$, of order $3840$, corresponds to the automorphism group of the code $((5,6,2))$, the first instance of a non-additive quantum code \cite{Rains}. 

\begin{table}[h]
\begin{center}
\begin{tabular}{|r|r|r|r||r|r|}
\hline
$g_i$& $g_2$ & $g_3$ & $g_4$ & $g_5$ & $g_6$ \\
\hline
$G$& $\mathcal{Z}_2^{\times 2}$ & $(\mathcal{Z}_4 \times \mathcal{Z}_2) \rtimes \mathcal{Z}_2$ & $ (\mathcal{Z}_2 \times \mathcal{Q}_8)\rtimes \mathcal{Z}_2$ & $\mathcal{Z}_2 \times((\mathcal{Z}_2 \times \mathcal{Q}_8)\rtimes \mathcal{Z}_2) $ & $g_6$\\
\hline
$\mbox{Aut}(G)$& $\mathcal{D}_8$ & $\mathcal{Z}_2 \times S_4$ & $\mathcal{Z}_2 \wr A_5$ & $\mathcal{Z}_2^{\times 2} \wr A_5$ & $\mathcal{Z}_2^{\times 3} \wr A_5$\\
\hline
$\left| \mbox{Aut}(G)\right|$& $8$ & $48$ & $1920$ & $61440$ & $1966080$ \\
\hline
\end{tabular}
\label{autom}
\caption{Group structure of an independant set of the two-qubit ($g_2$ to $g_4$) and three-qubit systems ($g_2$ to $g_6$). $G$ denotes the identified group and $\mbox{Aut}(G)$ the corresponding automorphism group. $\mathcal{Q}_8$ and $\mathcal{D}_8$ are the eight-element quaternion and dihedral groups.}
\end{center}
\end{table}

The same approach can be applied to the three-qubit system and higher-order qubit systems. For the three-qubit system, the size of a maximal independant set is found to be seven (it is different from the size $9=2^3+1$ of a complete set of MUBs). The corresponding automorphism group encompasses the one of the two-qubit system as shown in Table 1. The group $\mbox{Aut}(g_n)$ ($n>4$) is found to be isomorphic to the wreath product $\mathcal{Z}_2^{\times m} \wr A_5$, with $m=n-3$. Its central quotient equals its derived subgroup and may be identified to the perfect group $(\mathcal{Z}_2^{\times 4})^{\rtimes m}\rtimes A_5$. These perfect groups of order $960$, $15360$, $245760$ contain some elements, which are not commutators \footnote{The calculation is performed using theorem 6.6 in \cite{kappe}.}.

\section{Conclusion}

Advanced group theoretical tools may be used to explore fault tolerance in quantum computing. We found some fingerprints of quantum (de)coherence in exceptional groups such as $U_6$ (the stabilizer of an hexad in $M_{22}$), in the group $M_{20}$, and in the automorphism groups of mutually unbiased bases. Using this approach, disparate concepts such as the stabilizer formalism, topological quantum computing \cite{Bombin07} and the mathematical approach of quantum complementary, tend to merge. Future work will be devoted to arbitrary $n$-qudit systems and composite systems, and the link to quantum codes. 

\section*{Acknowledgements}
The authors acknowledge the support of the PEPS program (Projets Exploratoires Pluridisciplinaires) from the ST2I department at CNRS, France (Sciences et Technologies de l'Information et de la Communication).

\end{document}